\documentclass[aps,prab,reprint,groupedaddress]{revtex4-2}

\usepackage{siunitx}    
\DeclareSIUnit\atomicmassunit{u}

\usepackage{xcolor}     

\usepackage{tikz}
\usetikzlibrary{hobby,shapes.geometric,arrows.meta,decorations.markings,decorations.pathmorphing,external,snakes,external}
\usepackage{pgfplots}
\pgfplotsset{compat=1.15,/pgf/number format/1000 sep={},legend image code/.code={
    \draw[mark repeat=2,mark phase=2]
    plot coordinates {
    (0cm,0cm)
    (0.15cm,0cm)        
    (0.3cm,0cm)         
    };%
    },/pgfplots/colormap={Jet}{rgb(0cm)=(0,0,0.502) rgb(1cm)=(0,0,0.9) rgb(2cm)=(0,0.3,1) rgb(3cm)=(0,0.7,1) rgb(4cm)=(0.1,1,0.9) rgb(5cm)=(0.5,1,0.5) rgb(6cm)=(0.9,1,0.1) rgb(7cm)=(1,0.7,0) rgb(8cm)=(1,0.3,0) rgb(9cm)=(0.9,0,0) rgb(10cm)=(0.502,0,0)}}
\usepgfplotslibrary{patchplots,groupplots,fillbetween}

\usepackage{caption}
\usepackage{subcaption}

\begin{document}


\title{Machine learning-based virtual diagnostics of dielectric laser acceleration}

\author{T. Egenolf}
\email[]{thilo.egenolf@tu-darmstadt.de}
\author{O. Boine-Frankenheim}
\altaffiliation[Also at ]{GSI Helmholtzzentrum f\"ur Schwerionenforschung GmbH, Planckstrasse 1, 64291 Darmstadt, Germany}

\affiliation{Institute for Accelerator Science and Electromagnetic Fields (TEMF), Technische Universit\"at Darmstadt, Schlossgartenstrasse 8, 64289 Darmstadt, Germany}

\date{\today}

\begin{abstract}
We present the development of a digital twin–based reconstruction framework for dielectric laser acceleration (DLA) based on machine-learning-assisted inversion of single-shot electron energy spectra. DLA as a promising candidate for compact electron accelerator designs using optical near-fields in dielectric nanostructures lacks on direct diagnostics on the laser-electron interaction. Thus, the outgoing electron energy distribution is one of the few experimentally accessible observables. To exploit this information, the DLA interaction and the mapping on the downstream spectrometer are treated as nonlinear measurement device whose response is described by the symplectic six-dimensional tracking code DLAtrack6D. This forward simulation model serves as a digital twin mapping laser-electron interaction parameters onto the resulting energy spectrum.

For diagnostics, we are interested in the inverse mapping which is represented by a neural network trained by synthetic datasets generated with the forward simulation model. The reconstruction performance and parameter identifiability of the inverse model are evaluated for parameter ranges relevant to planned experiments at the ARES linac in the SINBAD facility at DESY. Simulation studies demonstrate that the method can recover pulse front tilt angles with an accuracy of about \SI{1}{\degree} and phase offsets with an RMSE of about \SI{0.36}{\radian} corresponding to a difference of \SI{0.4}{\femto\second} in arrival time. Training with noisy spectra further improves robustness against spectrometer noise.

The trained surrogate model evaluates in the millisecond range, enabling shot-to-shot parameter estimation compatible with the \SI{50}{\hertz} repetition rate of ARES. The approach effectively transforms the DLA interaction region into a virtual in situ diagnostics for otherwise inaccessible laser parameters during experiment and provides a foundation for real-time monitoring and control of future DLA experiments.\end{abstract}

\maketitle

\section{\label{sec:intro}Introduction}
Dielectric laser acceleration (DLA) has emerged as a promising approach toward compact accelerator technology by exploiting optical near-fields in nanostructured dielectric materials~\cite{England2014DielectricAccelerators,Peralta2013DemonstrationMicrostructure,Breuer2013Laser-basedStructure}. DLA structures are driven by infrared or optical laser pulses, enabling accelerating gradients in the multi-GV/m range due to the high damage thresholds of dielectric materials at optical frequencies~\cite{Peralta2013DemonstrationMicrostructure,Leedle2015LaserStructure}. Experimental demonstrations of laser-driven acceleration in nanophotonic structures have confirmed the feasibility of the concept at subrelativistic and relativistic electron beam energies \cite{Wootton2016,Cesar2018High-fieldAccelerator,Chlouba2023CoherentAccelerator,Broaddus2024SubrelativisticAccelerators}.

A characteristic feature of DLA structures is the sub-wavelength spatial periodicity following the synchronicity condition $\lambda_z=m\beta\lambda_0$, where $m$ is the order of the accelerating spatial harmonic, $\beta$ is the particles' velocity and $\lambda_0$ is the central laser wavelength~\cite{Niedermayer2017BeamScheme,Niedermayer2018Alternating-phaseAcceleration,Niedermayer2020ThreeAccelerators}. Typically, the period lengths are on the order of a few hundreds of nanometers up to a few micrometers and interaction lengths ranging from hundreds to thousands of grating periods. The accelerating fields are confined to near-field regions inside nanofabricated dielectric geometries, rendering direct field diagnostics during operation impractical. Unlike RF cavities, where field probes and pickup antennas provide access to amplitude and phase information, DLA structures do not allow internal electromagnetic measurements.

Consequently, the primary experimentally accessible observable in current DLA experiments is the outgoing electron energy spectrum measured downstream of the interaction region using a magnetic spectrometer~\cite{Mayet2018SimulationsSINBAD,Crisp2025DynamicPulses}. The spectrometer thus represents the only direct beam diagnostic that encodes information about the laser–electron interaction inside the nanophotonic structure on a shot-to-shot base during the experimental run. However, the mapping between laser parameters and the measured spectrum is highly nonlinear and depends on multiple spatio-temporal degrees of freedom, including pulse front tilt, spectral phase, arrival time offset, and transverse alignment.
Precise control of these parameters is essential for stable DLA operation. In particular, pulse front tilt (PFT) is required to extend interaction lengths~\cite{Cesar2018EnhancedLaser,Crisp2024ExtendedStructure}. Similarly, femtosecond-level timing synchronization between the electron bunch and the laser pulse critically determines the achievable net acceleration. Small deviations in shaping parameters can lead to significant reductions in effective accelerating gradient or to asymmetric spectral distortions~\cite{Crisp2025DynamicPulses}. The absence of direct internal diagnostics therefore poses a significant instrumentation challenge for DLA experiments.

In conventional accelerator facilities, beam-based alignment and RF phase scans are routinely used to infer cavity parameters indirectly. A similar philosophy can be adopted for DLA systems: the interaction region and downstream spectrometer can be interpreted as a nonlinear measurement device whose response function is determined by beam dynamics in the laser-driven near field. Under this perspective, the measured spectrum represents the output of a deterministic forward operator that maps a reduced set of effective laser parameters to an observable energy distribution.

The reconstruction of laser and interaction parameters from the measured spectrum constitutes an inverse problem. Due to the nonlinear dependence of the spectrum on shaping variables and the finite resolution of the spectrometer, the inversion is generally ill-posed in the full parameter space. However, when restricted to a physically motivated and experimentally bounded parameter domain, the problem becomes tractable.

In this work, we present the design and performance evaluation of a model-based virtual beam diagnostic for DLA experiments, e.g. at the ARES linac at DESY~\cite{Mayet2018SimulationsSINBAD,Dorda2018StatusDESY}. A fast six-dimensional tracking model is employed to compute the forward instrument response of a realistic DLA–spectrometer system~\cite{Niedermayer2017BeamScheme}. Based on this forward model, synthetic datasets covering the experimentally relevant parameter space are generated and used to train a neural network that approximates the inverse mapping from measured spectra to effective interaction parameters.
The proposed approach is not intended as a replacement for physical modeling but as a computationally efficient surrogate for iterative fitting procedures and shot-to-shot parameter estimations, enabling real-time monitoring of effective on-chip laser parameters. The method transforms the DLA interaction into a virtual diagnostic instrument capable of providing indirect access to otherwise inaccessible quantities.

The present paper focuses on the instrumentation aspects of this concept. We quantify the sensitivity of the spectrometer response to relevant shaping parameters, analyze identifiability and noise robustness, and evaluate the achievable reconstruction accuracy under realistic experimental conditions. While experimental validation is pending commissioning of the DLA laser setup at ARES, the results provide quantitative performance benchmarks and establish the methodological framework for real-time spectral inversion in nanophotonic accelerator experiments.

\begin{table}[htp]
    \caption{Simulation parameters according to experimental planes at ARES for a working point including velocity bunching~\cite{Mayet2018SimulationsSINBAD}.}
    \label{tab:parameters}
    \begin{ruledtabular}
        \begin{tabular}{clclc}
            &Parameter && Value&\\
            \colrule
            &Mean beam energy && \SI{99.1}{\mega\electronvolt} &\\
            &Initial rms energy spread && \SI{43.8}{\kilo\electronvolt}&\\
            &Initial bunch length && \SI{268}{\nano\metre}&\\
            &DLA structure length && \SI{1}{\milli\metre}&\\
            &DLA period length && \SI{2.05}{\micro\metre}&\\
            &Central laser wavelength && \SI{2.052}{\micro\metre}&\\
            &Acceleration gradient && \SI{1}{\giga\volt\per\metre}&\\
            &Repetition rate && \SI{50}{\hertz}&
        \end{tabular}
    \end{ruledtabular}
\end{table}

\section{\label{sec:forward}Forward Model and Instrument Response Definition}
To enable quantitative reconstruction of interaction parameters, the spectrometer response must be described as a deterministic forward operator that maps a reduced set of effective laser and alignment parameters onto the measured energy distribution. This requires a beam dynamics model that captures the essential features of the laser–electron interaction within the nanophotonic structure while remaining computationally efficient enough for large-scale dataset generation.
\begin{figure}[htp]
    \centering
    \includegraphics{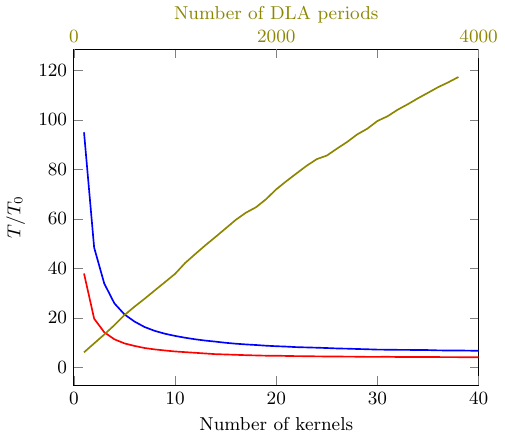}
    \caption{\label{fig:timing}Computing time of the Forward model using DLAtrack6D as function of the number of used CPU kernels (red and blue line for $10^5$ and $2.5\times10^5$ macroparticles, respectively) showing the parallelization properties of the applied tracking code, and as function of the number of simulated DLA periods showing the almost linear dependence.}
\end{figure}

In the following, we define the forward response function of the DLA–spectrometer system and specify the physically motivated parameterization used throughout this study. Particular emphasis is placed on numerical stability, computational performance, and consistency with realistic experimental operating conditions.

\begin{figure*}[htp]
    \begin{subfigure}[l]{0.45\textwidth}
        \includegraphics{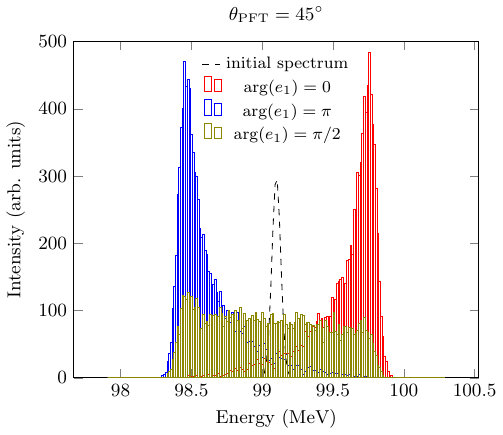}
    \end{subfigure}
    \hfill
    \begin{subfigure}[r]{0.45\textwidth}
        \includegraphics{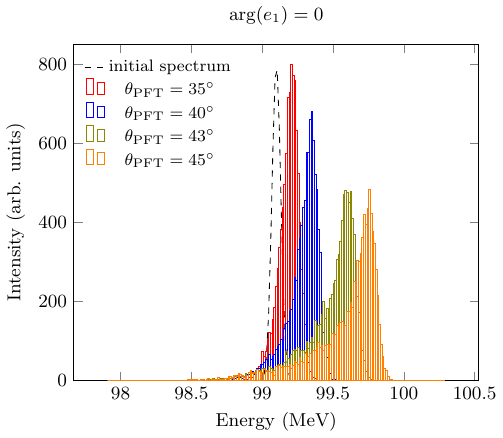}
    \end{subfigure}
    
    \caption{\label{fig:spectra}Simulated spectrometer response for different PFT angles and phases of the resonant harmonic. The initial Gaussian energy distribution (dashed line) evolves into a modulated spectrum due to the DLA interaction.}
\end{figure*}

To describe the DLA–spectrometer system as a measurement device, we introduce a reduced parameter vector $\theta\in\mathcal{R}^n$, where n is 3 to 5, typically. Possible parameters are the pulse front tilt angle $\theta_\mathrm{PFT}$, the phase offset encoded in $\mathrm{arg}(e_m)$, where 
\begin{equation}
    e_m(x,y)=\frac{1}{\lambda_z}\int_{-\lambda_z/2}^{\lambda_z/2}E_z(x,y,z)\exp{\left(im\frac{2\pi}{\lambda_z}z\right)}
\end{equation}
is the accelerating spatial harmonic of order $m$~\cite{Niedermayer2017BeamScheme}, the arrival time offset $\Delta t$ between laser pulse and electron bunch, transverse misalignment, the laser field amplitude and higher-order laser phase distortions. In the present study we focus on the pulse front tilt angle and the phase offset which also represents the arrival time jitter of the electrons. The investigated parameter ranges chosen according to realistic experimental planes for the ARES DLA experiment~\cite{Mayet2018SimulationsSINBAD} are shown in Tab.~\ref{tab:parameters}.

The arrival time jitter is assumed to be in the range of \SI{10}{\femto\second} (rms) which is larger than a DLA period length~\cite{Zhu2016Sub-fsChicane}. Thus, the simulated DLA interaction scans the complete $2\pi$ phase interval including accelerating and decelerating modes. Space-charge effects are neglected due to the low bunch charge regime (sub-pC) typical for DLA experiments. Radiation effects are negligible at the considered interaction lengths. 

The forward operator is evaluated using the six-dimensional symplectic tracking model DLAtrack6D~\cite{Niedermayer2017BeamScheme} that tracks macroparticles through the effective accelerating field of the DLA structure.
The longitudinal momentum update and transverse kicks per period by the synchronous Fourier harmonics of the accelerating field, which depends on the relative phase between electron and laser pulse and includes the effect of pulse front tilt by a tilted Gaussian intensity profile. For each parameter vector $\theta$, the model tracks typically $10^5$ macroparticles through the structure. The resulting energy distribution is histogrammed into 200 equidistant energy bins to represent the spectrometer output with a relative energy resolution of about $\Delta E/E\approx10^{-4}$. The computational time per forward evaluation scales linearly with the number of DLA periods and decreases exponentially with the number of used CPU cores (see Fig.~\ref{fig:timing}).

The complete instrument response can thus be expressed as the nonlinear mapping
\begin{equation}\label{eq:forward}
    \mathcal{F}:\theta\rightarrow S(E)\textrm{,}
\end{equation}
where $\mathcal{F}$ includes six-dimensional beam tracking inside the DLA structure, energy projection at structure exit, discretization into finite bins and a normalization to unit area before being used as input to the inverse reconstruction model.

As an example for forward model results of different input parameters, Fig.~\ref{fig:spectra} shows the simulated energy spectra for different input parameter sets. In the left plot, the phase is varied for a fixed pulse front tilt angle and in the right plot, the  PFT angle is varied for fixed phase, respectively. Both plots depict the highly non-linear evolution of the initially Gaussian energy spectrum as function of the input parameter vector $\theta$.

\begin{figure}[htbp]
    \centering
    \includegraphics{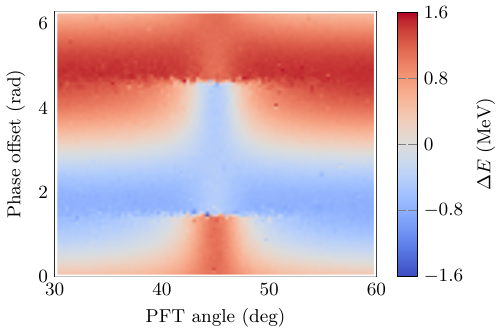}
    \caption{\label{fig:density}Simulated mean energy gain as a function of pulse front tilt angle and phase offset. The plot shows a quasi-symmetry plane in the PFT angle dependence at \SI{45}{\degree}.}
\end{figure}

\begin{figure}[htp]
    \centering
    \includegraphics{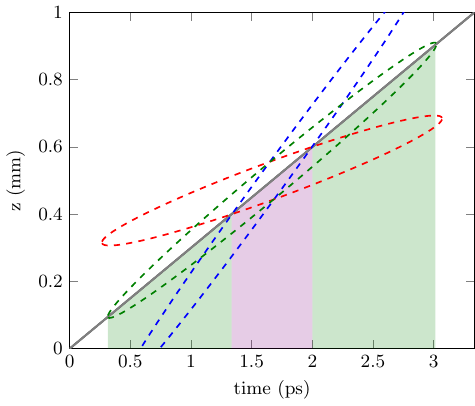}
    \caption{\label{fig:pft}Time and position of interaction between electron (black line) and laser pulse: longest interaction length for $\theta_\textrm{PFT,opt}=\SI{45}{\degree}$ (green ellipse), shorter interaction lengths for non-optimal PFT angles, but almost symmetric for $\theta_\textrm{PFT}=\theta_\textrm{PFT,opt}\pm\Delta\theta$ (red and blue ellipses).}
\end{figure}

The inversion problem addressed in the following section consists of approximating the restricted inverse operator
\begin{equation}\label{eq:inverse}
    \mathcal{F}^{-1}:S(E)\rightarrow\theta\textrm{,}
\end{equation}
using supervised learning on synthetic datasets generated by repeated evaluation of $\mathcal{F}$. To show that this inversion is well-defined at the studied parameter range, Fig.~\ref{fig:density} shows a head map of the mean energy gain as a function of both varied parameters. Although the inverse model is trained using the complete spectral information, the mean energy already provides some indications of the invertibility of the problem: along the axis of the phase offset the energy gain varies smoothly and a inversion should be feasible. However, along the axis of the pulse front tilt a symmetry plane is visible at $\theta_\textrm{PFT}\approx\SI{45}{\degree}$, i.e. at the optimal pulse front tilt angle for this relativistic electron energy. Fig.~\ref{fig:pft} also depicts the symmetry in interaction length between electrons and laser pulse. The resulting non-uniqueness of the reconstruction will be further discussed in Sec.~\ref{sec:performance}.

\section{\label{sec:inverse}Inverse Problem and Reconstruction Method}
Given the nonlinear forward operator $\mathcal{F}$ (see Eq.~\ref{eq:forward}), the diagnostic task consists of estimating the parameter vector $\theta$ from a measured spectrum $S_\textrm{meas}(E)$. Formally, the inverse problem can be written as $\textrm{find }\theta\textrm{ such that }\mathcal{F}(\theta)\approx S_\textrm{meas}(E)$ Due to the nonlinear, coupled dependence of the spectrum on the parameter vector $\theta$, the inverse problem is generally ill-posed in the full parameter space. Small perturbations in the measured spectrum, arising from detector noise or finite resolution, may lead to amplified parameter uncertainties depending on the local conditioning of the inverse operator.

However, when restricted to the bounded parameter set (see Tab.~\ref{tab:parameters}) and the two variable parameters pulse front tilt angle and phase offset introduced in Sec.~\ref{sec:forward}, the operator $\mathcal{F}$ remains locally smooth and exhibits sufficient sensitivity to the considered parameters. Under these conditions, a stable approximate inversion becomes feasible.

Direct iterative minimization of 
\begin{equation}
    ||\mathcal{F}(\theta)-S_\textrm{meas}||^2_2
\end{equation}
for each shot would be computationally prohibitive at the ARES repetition rate of \SI{50}{\hertz}, as it requires repeated forward model evaluations. Instead, we approximate the inverse mapping, Eq.~\ref{eq:inverse}, by training a neural network surrogate model
\begin{equation}
    \mathcal{G}_\phi(S)\approx\mathcal{F}^{-1}(S)\textrm{,}
\end{equation}
where $\phi$
 denotes the trainable network parameters. This approach can be interpreted as learning a digital twin as fast approximation to the solution of a nonlinear least-squares problem constrained to the bounded parameter domain.

 Training data are generated by random sampling of parameter vectors, followed by forward evaluation using the symplectic tracking code DLAtrack6D. The dataset in this study consists of 35000 training samples, 7000 validation samples and 5000 independent test samples. All input parameters and spectra are normalized prior to training. For robustness studies, additive Gaussian noise is applied to the spectra as
 \begin{equation}
    \label{eq:noise}
     S_\textrm{meas}(E)=S_\textrm{ideal}(E)+\epsilon(E)\textrm{,}\quad\epsilon(E)\propto\mathcal{N}(0,\sigma_\textrm{noise}^2)\textrm{.}
 \end{equation}

 The inverse surrogate model $\mathcal{G}_\phi$ is implemented using the machine learning framework TensorFlow~\cite{TensorFlowDevelopers2026TensorFlowv2.21.0} as fully connected feed-forward network with the 200 spectral energy bins as input dimensions, the two physical output parameters pulse front tilt angle and phase offset, two hidden layers with 200 neurons each and ReLU activation functions.
 The network is trained to minimize the mean absolute error loss
 \begin{equation}
     \mathcal{L}(\phi)=\frac{1}{N}\sum_{i=1}^N||\mathcal{G}_\phi(S_i)-\theta_i||_2\textrm{.}
 \end{equation}
 Training is performed using the Adam optimizer with early stopping based on validation loss.

 It is emphasized that the neural network does mot replace the physical model but serves as a computationally efficient approximation of the inverse response operator within the bounded parameter domain. The accuracy of the reconstruction is therefore fundamentally limited by the information content of the spectrum, the conditioning of the forward operator and the fidelity ot the physical model. The network architecture merely provides a fast nonlinear regression mechanism to approximate the inverse mapping.

\begin{figure*}[htp]
    \begin{subfigure}[l]{0.45\textwidth}
        \includegraphics{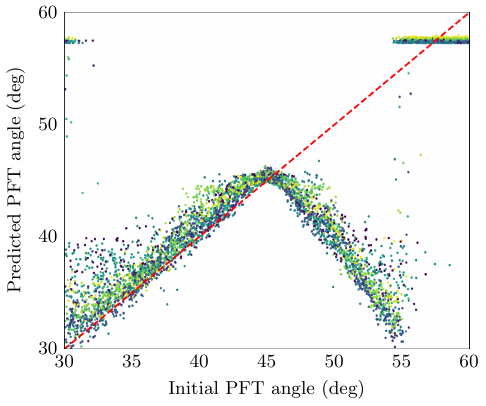}
    \end{subfigure}
    \hfill
    \begin{subfigure}[r]{0.45\textwidth}
        \includegraphics{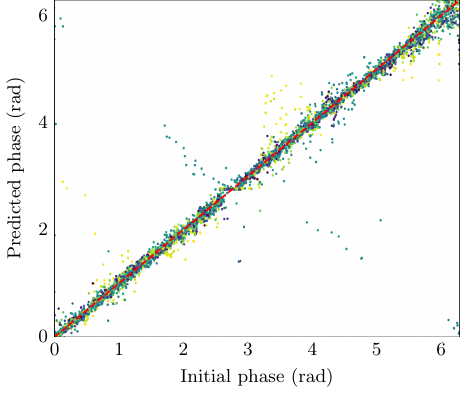}
    \end{subfigure}
    
    \caption{\label{fig:reconstruction}Reconstruction accuracy of pulse front tilt angle (left) and phase offset (right) on an independent test dataset (5000 samples). The neural network achieves a RMSE of about 1.5° in the PFT angle interval [\SI{35}{\degree},\SI{45}{\degree}] and a RMSE of 0.36 on the phase offset corresponding to \SI{0.4}{\femto\second} timing offset. The symmetry in the PFT angle is clearly visible. The plot color depicts the second parameter: \SI{30}{\degree} PFT angle, \SI{0}{\radian} phase offset (purple) $\rightarrow$ \SI{45}{\degree} PFT angle, $\pi$ phase offset (blue-green) $\rightarrow$ \SI{60}{\degree} PFT angle, $2\pi$ phase offset (yellow).}
\end{figure*}

 \section{\label{sec:performance}Reconstruction Performance and Robustness Analysis}
 The reconstruction performance is evaluated on the independent test dataset comprising 5000 parameter vectors $\theta=(\theta_\textrm{PFT},\textrm{arg}(e_1))$ randomly sampled from the bounded domain $\theta_\textrm{PFT}\in[\SI{30}{\degree},\SI{60}{\degree}]$ and $\textrm{arg}(e_1)\in[0,2\pi]$. None of these sampled ware used during training or validation. For each test sample $\theta_i$, the corresponding spectrum $S_i=\mathcal{F}(\theta_i)$ is generated using the forward model described in Sec.~\ref{sec:forward}. The training inverse model $\mathcal{G}_\phi$ then produces an prediction
 \begin{equation}
     \mathcal{\hat{\theta}_i}=\mathcal{G}_\phi(S_i)\textrm{.}
 \end{equation}
 Reconstruction accuracy is quantified using the root mean square error (RMSE) for each parameter component.

 Fig.~\ref{fig:reconstruction} shows the reconstructed pulse front tilt angle (left plot) and phase offset (right plot) as function of the ground-truth value, i.e. the initial parameter vector used in the forward model. The reconstruction error of the phase offset remains approximately uniform across the whole parameter range with an RMSE of \SI{0.36}{\radian}, corresponding to a timing offset of \SI{0.4}{\femto\second}. This indicates that the phase offset is well detectable under ideal noise-free conditions for the whole expected arrival time jitter range. Also, the pulse front tilt angle can be reconstructed with an RMSE of about \SI{1.5}{\degree} near below the optimal pulse front tilt angle of almost \SI{45}{\degree} with a slight increase near the boundary of the parameter range. However, the symmetry in the PFT angle dependence, already expected from the parameter study in Fig~\ref{fig:density}, is also clearly visible in the reconstruction accuracy. This leads to a swap of the reconstruction of angles above \SI{45}{\degree} to the wrong predictions. Introducing a second model to overcome this, is discussed in the following.

\begin{figure*}[htp]
    \begin{subfigure}[l]{0.45\textwidth}
        \includegraphics{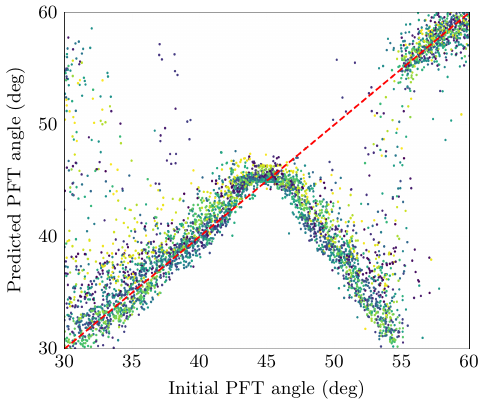}
    \end{subfigure}
    \hfill
    \begin{subfigure}[r]{0.45\textwidth}
        \includegraphics{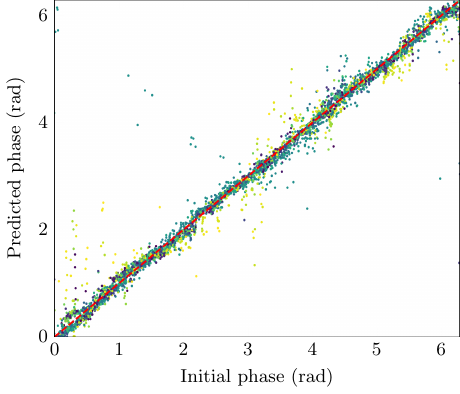}
    \end{subfigure}
    
    \caption{\label{fig:noisy_reconstruction}Reconstruction accuracy of pulse front tilt angle (left) and phase offset (right) on an independent test dataset (5000 samples). Gaussian noise is added to the training dataset of the neural network to improve the overall RMSE and the robustness against noisy measurement data for reconstruction. The plot color depicts the second parameter: \SI{30}{\degree} PFT angle, \SI{0}{\radian} phase offset (purple) $\rightarrow$ \SI{45}{\degree} PFT angle, $\pi$ phase offset (blue-green) $\rightarrow$ \SI{60}{\degree} PFT angle, $2\pi$ phase offset (yellow).}
\end{figure*}

However, before we discuss the split into two trained models, we take a look on the model trained by spectra with added Gaussian noise (see Eq.~\ref{eq:noise}). Since the ideal spectra used for the training are noise-free, it is expected that the reconstruction using the trained model is very sensitive to noisy measured spectral data. Thus, the model is trained with the same training and validation dataset as before, but adding Gaussian noise to the spectral data. Fig.~\ref{fig:noisy_reconstruction} shows again the reconstructed pulse front tilt angle and phase offset. Compared to the noise-free model shown in Fig.~\ref{fig:reconstruction}, the RMSE of the phase offset is slightly improved to \SI{0.31}{\radian}, but at the same time the RMSE of the pulse front tilt reconstruction is slightly increased. However, this is now tested with the ideal noise-free test dataset. 

\begin{figure*}[htp]
    \begin{subfigure}[l]{0.45\textwidth}
        \includegraphics{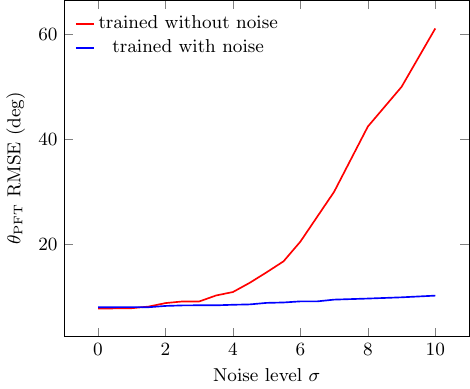}
    \end{subfigure}
    \hfill
    \begin{subfigure}[r]{0.45\textwidth}
        \includegraphics{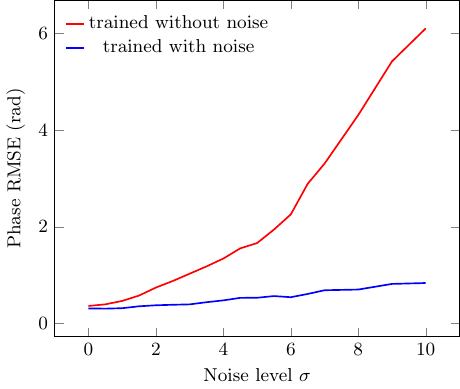}
    \end{subfigure}
    
    \caption{\label{fig:noise_rmse}Reconstruction error in pulse front tilt angle (left) and phase offset (right) as a function of Gaussian spectrometer noise level (see Eq.~\ref{eq:noise}). RMSE of the model trained on pure synthetic data increases very fast. However, performance remains stable for model trained with noisy data (blue lines) up to noise levels exceeding expected experimental conditions.}
\end{figure*}

Finally, adding random Gaussian noise also to the test dataset shows the improvement of the model trained with noisy spectral data. Fig.~\ref{fig:noise_rmse} shows the RMSE of pulse front tilt angle (left plot) and phase offset (right plot) for reconstructing the test dataset with different levels of the added Gaussian noise using the model trained by noise-free spectra and by the noisy spectra, respectively. Whereas the RMSE for noise-free training is already well above the acceptable limit, it remains within acceptable limits for training with noise across the entire noise level range of the test data.

\begin{figure*}[htp]
    \begin{subfigure}[l]{0.45\textwidth}
        \includegraphics{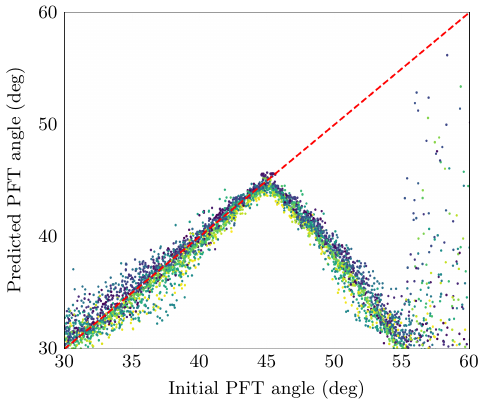}
    \end{subfigure}
    \hfill
    \begin{subfigure}[r]{0.45\textwidth}
        \includegraphics{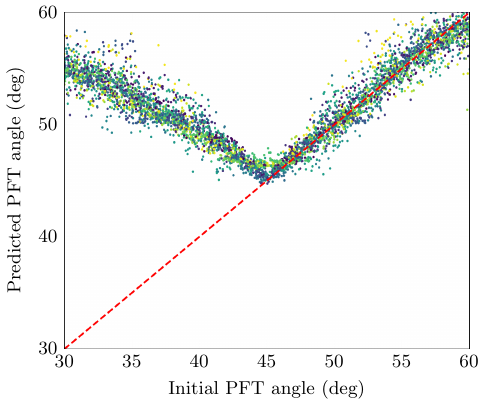}
    \end{subfigure}
    
    \caption{\label{fig:reconstruction_thetapft}Two models trained with pulse front tilt angles below \SI{45}{\degree} (left) and above \SI{45}{\degree} (right) offer the possibility to predict both possible PFT angles with improved RMSE. The plot color depicts the second parameter: \SI{30}{\degree} PFT angle, \SI{0}{\radian} phase offset (purple) $\rightarrow$ \SI{45}{\degree} PFT angle, $\pi$ phase offset (blue-green) $\rightarrow$ \SI{60}{\degree} PFT angle, $2\pi$ phase offset (yellow).}
\end{figure*}

\begin{figure*}[htp]
    \begin{subfigure}[l]{0.45\textwidth}
        \includegraphics{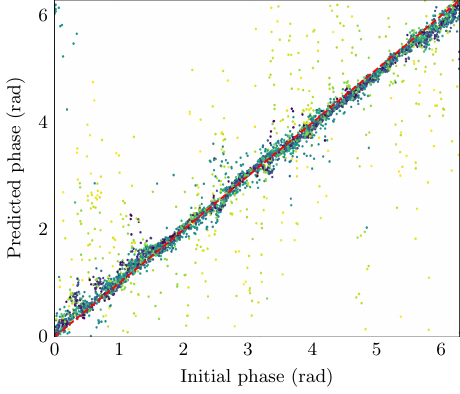}
    \end{subfigure}
    \hfill
    \begin{subfigure}[r]{0.45\textwidth}
        \includegraphics{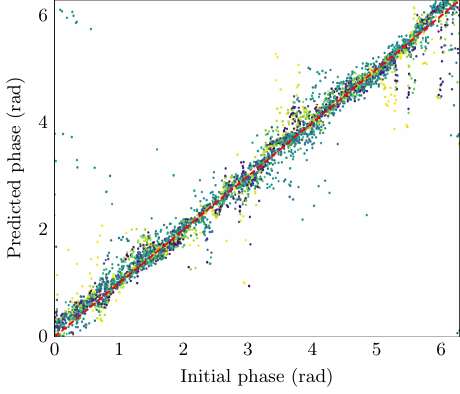}
    \end{subfigure}
    
    \caption{\label{fig:reconstruction_arge}The two models trained with pulse front tilt angles below \SI{45}{\degree} (left) and above \SI{45}{\degree} (right) can both predict the phase offset in the complete parameter range with similar accuracy. The plot color depicts the second parameter: \SI{30}{\degree} PFT angle, \SI{0}{\radian} phase offset (purple) $\rightarrow$ \SI{45}{\degree} PFT angle, $\pi$ phase offset (blue-green) $\rightarrow$ \SI{60}{\degree} PFT angle, $2\pi$ phase offset (yellow).}
\end{figure*}

To overcome the problem of non-unique reconstructions of the pulse front tilt angle due to the symmetry around $\theta_\textrm{PFT}\approx\SI{45}{\degree}$, the parameter range for training the inverse model is split into two separate domains above and below $\theta_\textrm{PFT}=\SI{45}{\degree}$, respectively. The reconstruction results of both models are shown in Figs~\ref{fig:reconstruction_thetapft} and~\ref{fig:reconstruction_arge}. Both plots indicate a clear improvement in the reconstruction accuracy for the trusted range below and above $\theta_\textrm{PFT}=\SI{45}{\degree}$, respectively. To utilize this improvement in the reconstruction framework, both models can be evaluated at each experimental shot and the trusted region can be chosen based on prior knowledge in a control loop, for example, or based on the reconstruction result of a subsequent shot with slightly changed pulse front tilt angle.

For application as a virtual beam diagnostic, the reconstruction must operate with the timing constraints imposed by the repetition rate of the used accelerator driving the experiment, i.e. \SI{50}{\hertz} at ARES. The forward evaluation of the operator $\mathcal{F}$ is above this timing constraints and a reconstruction or optimization requires many evaluations of the forward model. In contrast, the evaluation time of the trained inverse surrogate model is in the range of a millisecond, which enables a shot-to-shot evaluation corresponding to the \SI{20}{\milli\second} time budget per shot for a repetition rate up to \SI{50}{\hertz}.
Furthermore, the inverse framework enables the integration into a feedback loop by measuring the spectrum at the spectrometer screen, preprocess and normalize the measured data, reconstruct the parameter vector $\hat{\theta}$ and update the parameter settings for laser shaping independent of phase changes by arrival time jitter as shown above. This approach effectively transforms the DLA structure and spectrometer into a virtual in situ diagnostic for effective on-chip laser parameters, which are otherwise inaccessible due to the absence of diagnostics inside of the accelerating channel in the nanophotonic structures.

\section{\label{sec:conclusion}Conclusion}
In this work we presented a beam instrumentation concept for the reconstruction of complex spatio-temporal laser pulse structures from single-shot spectral measurements. The approach is motivated by the requirements of advanced dielectric laser acceleration (DLA) experiments, where the interaction between ultrafast electron bunches and optical near fields depends sensitively on the detailed spatio-temporal structure of the driving laser pulse.

A forward model describing the propagation of structured laser pulses through the interaction region and the spectrometer response was developed further to relate the measured spectrum to the underlying spatio-temporal electric field distribution. Numerical studies show that these pulse structures lead to characteristic and distinguishable spectral signatures after DLA interaction and propagation through the diagnostic system.
Based on this forward description, the reconstruction of the underlying pulse parameters was formulated as an inverse problem. A machine-learning based approach using neural networks was proposed to infer the relevant pulse parameters directly from measured spectra. Synthetic training data generated from the forward model allow the network to learn the mapping between spectral features and the corresponding spatio-temporal pulse parameters. Simulation studies indicate that the proposed method can recover key pulse-shaping parameters with good accuracy within realistic instrumental constraints.
The achievable reconstruction performance was evaluated under experimentally relevant conditions, including spectrometer resolution, finite spectral range, and measurement noise. The results indicate that the parameter space relevant for planned experiments at ARES at DESY can be resolved within the expected diagnostic capabilities.

While the present work focuses on simulations due to the delayed availability of experimental data, the framework provides a quantitative basis for the design and optimization of a virtual diagnostic system based on DLA interaction. In particular, the analysis highlights which spectral features contain the most information about the underlying pulse structure and therefore guides both the spectrometer configuration and the training of reconstruction algorithms.
The proposed approach combines physically motivated forward modeling with data-driven inversion and can therefore serve as a flexible diagnostic tool for complex ultrafast laser fields. Beyond the immediate application to DLA experiments, the method may also be relevant for other beam instrumentation tasks in ultrafast accelerator and laser–matter interaction experiments where the characterization of structured optical pulses is required.

\section{\label{sec:outlook}Outlook}
The simulation studies presented in this work provide the basis for the implementation of the proposed diagnostic concept in the upcoming experimental program at ARES at the SINBAD facility at DESY. In the next stage, the spectrometer-based diagnostic will be integrated into the laser–electron interaction beamline, where it will be used to record single-shot spectra of structured ultrafast laser pulses under realistic operating conditions. This will allow a first experimental validation of the forward model and the neural-network-based reconstruction approach.
A key objective of the initial experiments will be the systematic verification of the sensitivity of the measured spectra to controlled pulse-shaping parameters, such as pulse-front tilt and spatial chirp. For this purpose, programmable pulse-shaping elements will be used to generate well-defined spatio-temporal pulse structures within the shaping range considered in the simulations~\cite{Genovese2025LaserAcceleration}. The comparison between reconstructed and independently controlled pulse parameters will provide a quantitative benchmark of the reconstruction accuracy.
In the longer term, the presented approach may enable real-time monitoring and optimization of complex laser pulse structures in accelerator experiments. As mentioned before, such capabilities would be particularly valuable for dielectric laser acceleration setups, where the electron–laser interaction is highly sensitive to the spatio-temporal structure of the driving optical field.

\section*{Data availability}
The spectral data to train, validate and test the neural network proposed in this article are openly available~\cite{Egenolf2026Dataset:10.48328/tudatalib-2183}. The tracking code DLAtrack6D is available upon reasonable request from the authors.

\begin{acknowledgments}
The authors would like to thank H. Cankaya and L. Genovese (UHH/DESY) and the team of ARES at DESY for the opportunity to join the DLA beamtimes and very fruitful discussions. This work is funded by the German Federal Ministry of Research, Technology and Space (Grant Nos. FKZ: 05K22RDC and 05K25RD1).
\end{acknowledgments}

\bibliography{references}

\end{document}